\documentclass[twocolumn,showpacs]{revtex4}
\usepackage[T1]{fontenc}
\usepackage[latin1]{inputenc}
\usepackage{amsmath}
\usepackage{amssymb}
\usepackage{wasysym}
\usepackage[dvips]{graphicx}
\usepackage{esint}

\makeatletter
\@ifundefined{textcolor}{}
{%
 \definecolor{BLACK}{gray}{0}
 \definecolor{WHITE}{gray}{1}
 \definecolor{RED}{rgb}{1,0,0}
 \definecolor{GREEN}{rgb}{0,1,0}
 \definecolor{BLUE}{rgb}{0,0,1}
 \definecolor{CYAN}{cmyk}{1,0,0,0}
 \definecolor{MAGENTA}{cmyk}{0,1,0,0}
 \definecolor{YELLOW}{cmyk}{0,0,1,0}
}

\usepackage{graphicx}
\usepackage{amsfonts}
\def\be{\begin{equation}}
\def\ee{\end{equation}}
\def\bea{\begin{eqnarray}}
\def\eea{\end{eqnarray}}

\makeatother

\begin{document}

\title{Biased doped silicene as a source for advanced electronics}

\author{Y.G. Pogorelov,$^{1}$ V.M. Loktev$^{2}$ }

\affiliation{$^{1}$IFIMUP-IN, Departamento de F\'{i}sica, Universidade do Porto,
Portugal,\\
 $^{2}$Bogolyubov Institute for Theoretical Physics, National Academy
of Sciences of Ukraine, Kyiv, Ukraine; National Technical University
of Ukraine \textquotedbl{}Kyiv Polytechnic Institute\textquotedbl{},
Kyiv, Ukraine}
\begin{abstract}
Restructuring of electronic spectrum in a buckled silicene monolayer
under some applied voltage between its two sublattices and in presence
of certain impurity atoms is considered. A special attention is given
to formation of localized impurity levels within the band gap and
the to their collectivization at finite impurity concentration. It
is shown that a qualitative restructuring of quasiparticle spectrum
within the initial band gap and then specific metal-insulator phase
transitions are possible for such disordered system and can be effectively
controlled by variation of the electric field bias at given impurity
perturbation potential and concentration. Since these effects are
expected at low impurity concentrations but at not too low temperatures,
they can be promising for practical applications in nanoelectronics
devices. 
\end{abstract}


\maketitle

\section{Introduction}

\label{sec:Intro}After revolutionary breakthrough of graphene, introducing
unusual relativistic effects into solid state physics \citep{Geim},
the family of relative materials is continuously growing. In particular,
a possibility of obtaining new semiconducting materials where the
bandgap can be tuned by external electric bias is extensively studied.
This was first demonstrated for the graphene bilayer, called bigraphene
\citep{cast}. Unlike the basic monolayer graphene, here a non-equivalence
of two sublayers takes place under electric field applied normally
to them. Opening of a tunable semiconducting gap and perspectives
of its practical use in tunable transistors is now broadly discussed.
The same possibility was already indicated in a variety of similar
systems, including even single layered, as, for instance, silicene,
the Si-based analog of graphene \citep{Guzman}. Its important structural
difference from graphene consists in a much more pronounced buckling
of its 2D hexagonal lattice, thus leading to a non-equivalence of
two sublattices in the same layer and to opening of a bandgap under
normal-to-plane electric bias\citep{Ni,Quhe,Stille,Zand}. Subsequently,
fabrication of practical field-effect transistors based on a silicene
sheet is extensively sought for \citep{Drum,Li,Zand}.

When comparing these 2D systems with common semiconductors, an important
question arises on their behavior under doping by impurity atoms.
As well known, such doping in common semiconductors produces localized
in-gap energy levels near the edges of conduction band (called donor
levels) or valence band (acceptor levels) \citep{Shklo}. For mostly
used dopants (as Si neighbors from the periodic table) these levels
are very shallow (of some tens meV depth compared to some eV bandwidths)
so that charge carriers can be thermally excited from them to the
nearby band (conduction or valence) and thus contribute into conduction
of respective kind (electron or hole). The resulting conductivity
turns sensitive to external bias realized in specific devices, defining
their effectiveness \citep{Sze}. Typical dopant concentration $n$
is quite low, in order to assure the mean distance $\bar{r}=n^{-1/3}$
between dopants to surpass the long effective radius of localized
state $r_{loc}\gg a$ (the lattice parameter), it should not exceed
$n_{0}=r_{loc}^{-3}\sim10^{-17}$ cm$^{-3}$ (for 3D systems). Then
it is known that for $n\gg n_{0}$ the doped system is brought to
metallization \citep{Roth}, due to growing interaction between dopants
and subsequent broadening of the dopant level. Hence the Fermi level,
initially fixed at the dopant level, becomes displaced to the band
interior. This phenomenon is generally considered adverse for electronics
purposes, since it drastically reduces the bias sensitivity. Otherwise,
an alternative type of impurities (as transition and rare-earth elements),
producing the so called deep levels in the semiconducting gap \citep{Watkins},
are not effective for thermalization of carriers and instead can act
as traps for them.

The above limitations however can be effectively overcome under the
possibility for tuning the fundamental bandgap and also other relevant
spectrum characteristics as the Fermi level and the Mott's mobility
edges \citep{mott}. This opens a formerly unexplored perspective
of bringing the bias sensitivity of conduction to a much broader scale
then in common semiconductors, spanning it from metallic to insulating
regimes through the Mott's metal-insulator transition (MIT). The purpose
of the present study is to illustrate such an expectation on the particular
example of biased and doped silicene, considering both situations
of shallow and deep dopant levels with their specific regimes. Though
a detailed treatment of these issues, using the realistic impurity
potentials, their screening by relativistic electrons, etc., can present
certain technical problems, it can be much facilitated with use of
simplified models, traditionally applied for impurities both in common
semiconductors and in graphene-related materials. Such are the Lifshitz
model \citep{lif}, better suited for shallow dopants, and the Anderson
hybrid model \citep{and}, more adequate for deep dopants. Below we
analyze the electronic spectra of biased and doped silicene within
these two models and indicate possible tuning regimes to reach desirable
effects.

\section{Formulation of the problem}

\label{sec:Form}For the silicene hexagonal lattice with two non-equivalent
sites in unit cell (Fig. \ref{eq:1}), we write down the tight-binding
Hamiltonian as: 
\begin{equation}
H_{0}=\sum_{\mathbf{k}}\psi_{\mathbf{k}}^{\dagger}\hat{h}{}_{{\bf k}}\psi_{\mathbf{k}}.\label{eq:1}
\end{equation}
Here 2-spinors $\psi_{\mathbf{k}}^{\dagger}=\left(a_{{\bf k}}^{\dagger},b_{{\bf k}}^{\dagger}\right)$
are made of 2D Fourier transforms $a_{{\bf k}}=N^{-1/2}\sum_{{\bf n}}{\rm e}^{i{\bf k}\cdot{\bf n}}a_{{\bf n}}$
and $b_{{\bf k}}=N^{-1/2}\sum_{{\bf n}}{\rm e}^{i{\bf k}\cdot{\bf n}}b_{{\bf n}}$
of local Fermi operators at A- and B-type sites in the ${\bf n}$-th
unit cell ($N$ the total number of cells). The 2$\times$2 Pauli
matrix expansion $\hat{h}_{{\bf k}}=\hat{\sigma}_{3}V/2+\hat{\sigma}_{+}t_{{\bf k}}+\hat{\sigma}_{-}t_{{\bf k}}^{\ast}$
includes the on-site energy shifts $\pm V/2$, due to the effect of
buckling and external electric field, referred to as bias in what
follows, and the complex factors $t_{{\bf k}}=t\sum_{\boldsymbol{\delta}}{\rm e}^{i{\bf k}\cdot\boldsymbol{\delta}}$,
due to the hopping amplitude $t\sim1.1$ eV between nearest neighbor
sites separated by the vectors $\left(\boldsymbol{\delta},d_{z}\right)$
with their $xy$-plane components $|\boldsymbol{\delta}|=a\approx0.22$
nm and normal buckling components $d_{z}\sim0.2a$ \citep{Cahan,Jose}.
For sake of simplicity, relatively weak spin-orbit interactions and
spin degrees of freedom are omitted in Eq. \ref{eq:1}. The relevant
low-energy physics is generated near the nodal points $\pm{\bf K}=(\pm4\pi/3\sqrt{3}a,0)$
in the Brillouin zone so that for ${\bf q}={\bf k}-{\bf K}$ with
$aq\ll1$ we have $t_{{\bf k}}\approx\hbar v_{{\rm F}}q{\rm e}^{-i\varphi_{{\bf q}}}$
where the Fermi velocity $v_{{\rm F}}=3at/(2\hbar)$ and $\varphi_{{\bf q}}=\arctan q_{y}/q_{x}$. 

\begin{figure}
\includegraphics[bb=60bp 170bp 880bp 600bp,clip,scale=0.3]{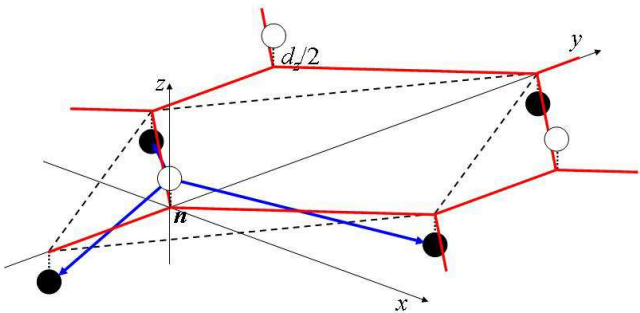}
\protect\caption{\label{Fig1}Crystalline structure of a buckled silicene plane where
silicon atoms are up- or down-shifted by $d_{z}/2$ from the initial
hexagonal plane (red lines). The dashed line delimits a unit cell
at the position ${\bf n}$ with two non-equivalent sites, of A-type
(clear) and B-type (dark), blue arrows indicate three vectors $\boldsymbol{\delta}$
between nearest neighbor Si atoms.}
\end{figure}

We study electronic states in this system using the Green function
(GF) matrix $\hat{G}(\mathbf{k},\mathbf{k}')=\langle\langle\psi_{\mathbf{k}}|\psi_{\mathbf{k'}}^{\dagger}\rangle\rangle$that
satisfies the equation of motion: 
\begin{equation}
\varepsilon\hat{G}(\mathbf{k},\mathbf{k}')=\langle\left\{ \psi_{\mathbf{k}},\psi_{\mathbf{k'}}^{\dagger}\right\} \rangle+\langle\langle\left[\psi_{\mathbf{k}},H\right]|\psi_{\mathbf{k'}}^{\dagger}\rangle\rangle,\label{eq:2}
\end{equation}
for the full Hamiltonian $H$. \label{Fig1-1}Generally, the GF matrix
defines the system energy spectrum by the roots of the general dispersion
equation: $\mathrm{Re\, det}\hat{G}=0$, that is by the poles of its
diagonal elements in the basis of exact eigen-states, and the total
density of states (DOS) is defined as:
\begin{equation}
\rho\left(\varepsilon\right)=\frac{1}{2\pi N}\mathrm{Tr\, Im}\,\hat{G},\label{eq:3}
\end{equation}
in any basis. Using of Eq. \ref{eq:2} with the unperturbed Hamiltonian,
$H=H_{0}$, Eq. \ref{eq:1}, leads to a momentum-diagonal form for
the non-perturbed GF matrix $\hat{G}^{\left(0\right)}({\bf k},{\bf k}')=\delta_{{\bf k},{\bf k'}}\hat{G}_{{\bf k}}^{\left(0\right)}$
where:
\begin{equation}
\hat{G}_{{\bf k}}^{\left(0\right)}=\frac{\varepsilon+\hat{h}_{{\bf k}}}{\varepsilon^{2}-V^{2}/4-\xi^{2}},\label{eq:4}
\end{equation}
and $\xi_{{\bf k}}=|t_{\mathbf{k}}|$ is an effective momentum variable
for the relevant low-energy range, $|\varepsilon|\ll\Lambda$ (where
$\Lambda=\hbar v_{\mathrm{F}}\sqrt{K/a}=t\sqrt{\pi\sqrt{3}}$ is the
bandwidth, that is an energy cut-off). In this approximation, the
electronic spectrum consists of two symmetric bands $\pm\varepsilon_{0}\left(\xi\right)=\pm\sqrt{V^{2}/4+\xi^{2}}$
with a gap of $V$ between them. The related non-perturbed DOS is
linear beyond this gap: 
\begin{equation}
\rho_{0}\left(\varepsilon\right)=\frac{\varepsilon}{\Lambda^{2}}\theta\left(V^{2}/4-\varepsilon^{2}\right)\theta\left(\Lambda^{2}+V^{2}/4-\varepsilon^{2}\right),\label{eq:5}
\end{equation}
and normalized: $\int_{-\infty}^{\infty}\rho_{0}\left(\varepsilon\right)d\varepsilon=1$.
This situation in silicene with buckled lattice is referred to as
realization of a tunable gap between the lower (valence) band and
the higher (conduction) band, in contrast to a fixed gap in common
semiconductors (like crystalline Si). 

Now we pass to the specifics of doping this system by impurity atoms
at random sites within its crystalline lattice that produce its perturbation
through the above referred Lifshitz and Anderson models.

\section{Lifshitz model}

\label{sec:Lif}We begin with the simpler Lifshitz model \citep{lif}
characterized by a single perturbation parameter, the on-site energy
shift $U$ on all impurity atoms located at ${\bf p}_{1}$ sites in
the 1st sublattice and ${\bf {\bf p}}_{2}$ sites in the 2nd sublattice
with relative concentrations $c_{1,2}$ (both expectedly small, $c_{j}\ll1$).
Such substitutional type better corresponds to impurities such as
Si neighbors in the periodic table. The corresponding Lifshitz perturbation
reads:
\begin{equation}
H_{L}=\frac{1}{N}\sum_{\mathbf{k},\mathbf{k}'}\sum_{j,{\bf p}_{j}}{\rm e}^{i\left({\bf k}'-{\bf k}\right)\cdot{\bf p}_{j}}\psi_{\mathbf{k}}^{\dagger}\hat{U}_{j}\psi_{\mathbf{k'}},ds\label{eq:6}
\end{equation}
with the scattering matrices $\hat{U}_{j}=U\hat{p}_{j}$ and the sublattice
projectors $\hat{p}_{1,2}=\left(1\pm\hat{\sigma}_{z}\right)/2$. In
this model with the full Hamiltonian $H=H_{0}+H_{L}$, the explicit
equation of motion for the momentum-diagonal GF matrix:
\begin{eqnarray}
\hat{G}(\mathbf{k}) & = & \hat{G}^{\left(0\right)}({\bf k})\nonumber \\
 & + & \frac{1}{N}\sum_{j,{\bf p}_{j},{\bf k}'}{\rm e}^{i\left({\bf k}'-{\bf k}\right)\cdot{\bf p}_{j}}\hat{G}_{{\bf k}}^{\left(0\right)}\hat{U}_{j}\hat{G}(\mathbf{k}',\mathbf{k}),\label{eq:7}
\end{eqnarray}
leads to the standard solution:

\begin{equation}
\hat{G}_{\mathbf{k}}=\left[\left(\hat{G}_{{\bf k}}^{\left(0\right)}\right)^{-1}-\varSigma_{1,{\bf k}}\hat{p}_{1}-\varSigma_{2,{\bf k}}\hat{p}_{2}\right]^{-1}.\label{eq:8}
\end{equation}
Here the partial self-energy functions are presented by their respective
group expansions (GE's) \citep{iva}: 
\begin{equation}
\varSigma_{j,{\bf k}}=c_{j}T_{j}\left(1+c_{j}B_{j,{\bf k}}+\cdots\right),\label{eq:9}
\end{equation}
where the partial T-matrix $T_{j}=U/\left(1-Ug_{j}\right)$ with local
GF's $g_{j}=N^{-1}\sum_{{\bf k}}\left(\hat{G}_{\mathbf{k}}\right)_{jj}$
describes the effects of multiple scatterings on single impurity center.
It can be generally shown that the GE series, Eq. \ref{eq:9}, is
converging within the energy range of band-like states and can be
well approximated there by its first T-matrix term while the rest
of terms are important for the check of convergence. The first of
non-trivial GE terms is due to impurity pairs:
\begin{equation}
B_{j,{\bf k}}=\sum_{{\bf n}\neq0}\frac{A_{j,{\bf n}}\mathrm{e}^{-i{\bf k}\cdot{\bf n}}+A_{j,{\bf n}}A_{j,-{\bf n}}}{1-A_{j,{\bf n}}A_{j,-{\bf n}}},\label{eq:10}
\end{equation}
and includes the functions $A_{j,{\bf n}}=T_{j}N^{-1}\sum_{{\bf k'\neq\mathbf{k}}}\mathrm{e}^{i{\bf k'}\cdot{\bf n}}\left(\hat{G}_{\mathbf{k'}}\right)_{jj}$
of inter-impurity interaction. It should be also noted that all the
products of these functions in the expansion of Eq. \ref{eq:10} are
presented by multiple sums in \textit{non-coinciding} momenta \citep{iva,ILP}.
The omitted terms in the brackets of Eq. \ref{eq:9} correspond to
clusters of three and more impurities, they are expressed through
respective combinations of these functions. 

Generally, once the initial translation symmetry of crystalline lattice
is broken by the presence of impurities, the quasi-momentum is no
longer an exact quantum number and also the system spectrum is not
limited to the initial bands, since already a single impurity can
produce localized levels beyond the bands. However, for not too strong
a disorder, this spectrum maintains continuous ranges of band-like
states (both the modified initial bands and possibly some new, impurity,
bands arisen near localized levels), intercalated by the ranges of
truly localized states (either on single impurities or on their clusters).
Within band-like ranges, the approximate dispersion laws of corresponding
subbands $\varepsilon_{j}\left(\xi\right)$ (in our case, $j=1,2,imp$
for two initial and impurity bands respectively) are given by the
formal roots of the above mentioned dispersion equation in the $\mathbf{k}$-basis:
\begin{equation}
\mathrm{Re\, det}\,\hat{G}_{\mathbf{k}}=0.\label{eq:11}
\end{equation}
However their validity is restricted by the known Ioffe-Regel-Mott
(IRM) criterion \citep{ir,mott} that for this case takes the form:
\begin{equation}
\xi d\varepsilon_{j}\left(\xi\right)/d\xi\gg\Gamma_{j}\left(\xi\right)\label{eq:12}
\end{equation}
with the damping term $\Gamma_{j}\left(\xi\right)=\mathrm{Im}\varSigma_{j,{\bf k}}$
for $\xi=\xi_{{\bf k}}$. The separation points between extended and
localized ranges, called Mott's mobility edges, are estimated from
the condition that the symbol ``$\ll$'' in Eq. \ref{eq:12} is
changed for ``$\sim$'', ending validity of the IRM criterion. This
also qualitatively agrees with a similar change in the convergence
criterion for GE: $c|B_{j,{\bf k}}|\ll1$. Within the bandgap, the
broadening $\Gamma_{j}$ mainly results from $\mathrm{Im}\, B_{j,{\bf k}}$
in Eq. \ref{eq:9}, so analysis of this range needs calculation of
the functions $g_{j}$ and $A_{j,{\bf n}}$. A reasonable approximation
for them follows from substitution of $\hat{G}_{\mathbf{k}}$ by $\hat{G}_{{\bf k}}^{\left(0\right)}$
in corresponding sums resulting in: 
\begin{equation}
g_{1,2}\approx\frac{\varepsilon\pm V/2}{W^{2}}\ln\frac{V^{2}/4-\varepsilon^{2}}{\Lambda^{2}}\label{eq:13}
\end{equation}
and:

\begin{equation}
A_{j,{\bf n}}\approx\frac{\left(\varepsilon\pm V/2\right)T_{j}}{\Lambda^{2}}K_{0}\left(n/r_{\varepsilon}\right)\label{eq:14}
\end{equation}
(see details in Appendix). Here the characteristic length $r_{\varepsilon}=\hbar v_{\mathrm{F}}/\sqrt{V^{2}/4-\varepsilon^{2}}$
and the McDonald function $K_{0}\left(x\right)$ has asymptotics \citep{Abst}:
\[
K_{0}\left(x\right)\approx\begin{cases}
\ln\left(2/x\right)-\gamma, & x\ll1,\\
\sqrt{2/\left(\pi x\right)}\mathrm{e^{^{-x}}}, & x\gg1,
\end{cases}
\]
with the Euler's constant $\gamma\approx0.5772$. 

The logarithmic divergence of $g_{j}$, Eq. \ref{eq:13}, near one
of the gap edges allows a localized level $\varepsilon_{loc}$ to
appear there under a proper impurity perturbation. Thus, if one chooses
for definiteness $U<0$ (and supposedly $|U|\lesssim\Lambda$), this
level is due to the pole of $T_{1}$ (by impurities in 1st sublattice)
near the upper edge $V/2$, their separation being well approximated
as:
\begin{equation}
V/2-\varepsilon_{loc}\approx\frac{\Lambda^{2}}{V}\mathrm{e}^{-\Lambda^{2}/\left(|U|V\right)}\equiv c_{0}\frac{\Lambda^{2}}{V}.\label{eq:15}
\end{equation}
Hence the localized level is exponentially shallow for all practically
achievable bias values (always $V\ll\Lambda$), which can justify
such modeling of real shallow levels. At this choice, another term
$T_{2}$ (by impurities in 2nd sublattice) has no poles and is less
relevant. 

Further analytical study of modified spectrum uses some approximated
energy dependencies of the relevant $T_{1}$-matrix. Thus, in a close
enough vicinity to the localized level, $|\varepsilon-\varepsilon_{loc}|\ll V/2-\varepsilon_{loc}$,
its denominator can be linearized:
\begin{equation}
T_{1}\approx\frac{\Lambda^{2}\left(V/2-\varepsilon_{loc}\right)}{V\left(\varepsilon-\varepsilon_{loc}\right)},\label{eq:16}
\end{equation}
while in a wider area, $0<V/2-\varepsilon\ll V$, the logarithmic
approximation applies:
\begin{equation}
T_{1}\approx\frac{\Lambda^{2}}{V}\ln^{-1}\frac{V/2-\varepsilon_{loc}}{V/2-\varepsilon}.\label{eq:17}
\end{equation}
At last, when considering the energy scales over the whole band gap,
$|\varepsilon|\apprge V$, the complete formula, Eq. \ref{eq:13},
should be used in the T-matrices. 

\begin{widetext}

\begin{figure}[t]
\includegraphics[bb=80bp 80bp 920bp 650bp,clip,scale=0.6]{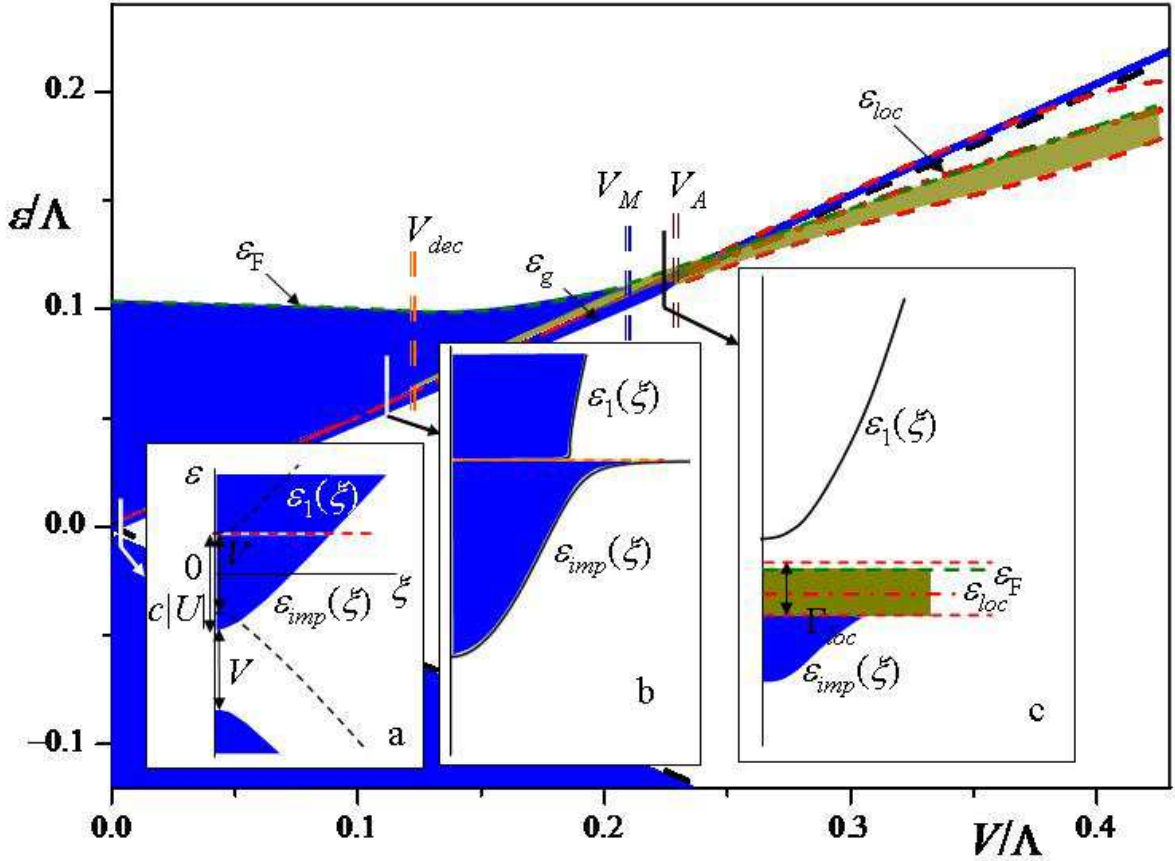}

\protect\caption{\label{Fig2}Spectrum restructuring in silicene with Lifshitz impurities
at the choice of their parameters $U=-\Lambda$ and $c=0.01$ in function
of the bias $V$. Insets show the particular dispersion laws at: a)
weak ($V=0.004\Lambda$), b) medium ($V=0.11\Lambda$), and c) strong
($V=0.23\Lambda$) bias. Blue areas present filled band-like states
and dark yellow do filled localized states, red dashed lines mark
the concentrational broadening of impurity level. Double dashed lines
indicate the critical bias levels $V_{dec}$ for band decomposition
(orange), $V_{M}$ for Mott (blue), and $V_{A}$ Anderson (purple)
phase transitions.}
 
\end{figure}

\end{widetext}

Of course, for $U>0$, symmetric formulas with respect to Eqs. \ref{eq:15},
\ref{eq:16}, \ref{eq:17} apply for the pole of $T_{2}$ near the
lower gap edge and for its related vicinities while $T_{1}$ becomes
irrelevant. 

Let us focus now on the most restructured region of spectrum, including
the impurity band $\varepsilon_{imp}(\xi)$ and its closest neighbor
areas of $\varepsilon_{1,2}(\xi)$ bands. In this course, it is convenient
to consider this restructuring with growing bias $V$ at fixed impurity
parameters $U$ and $c$. The T-matrix approximation$\varSigma_{j,{\bf k}}\approx c_{j}T_{j}$
is sufficient at the first step, as far as the quasiparticle lifetime
and respective IRM limits for band-like states are not considered.

At lowest bias $V\ll c|U|$, the numerical solution of the dispersion
equation, Eq. \ref{eq:11}, with use of Eq. \ref{eq:13}, shows the
spectrum restructuring to be very close to its simple shift by $cU$,
that is the impurity effect is reduced to that of effective medium
potential. Formally, this solution includes the impurity band $\varepsilon_{imp}(\xi)$
of $c|U|$ width and separately the modified upper subband $\varepsilon_{1}(\xi)$.
However, it is seen from Fig. \ref{Fig2}(inset a) that their composition
is closely matched near the $\varepsilon_{loc}$ level and practically
coincides with a single shifted law: $\varepsilon_{comp}\left(\xi\right)=\sqrt{V^{2}/4+\xi^{2}}+cU$,
attaining its lower edge at $\xi=0$: $\varepsilon_{g}\equiv\varepsilon_{comp}(0)=V/2+cU$.
As seen from Fig. \ref{Fig2}(inset b), such composite $\varepsilon_{1}+\varepsilon_{imp}$-band
structure persists for low enough bias such that $c\ll c_{0}$, and,
using Eq. \ref{eq:15}, this relates to: $V\ll V_{A}=\Lambda^{2}/\left(|U\ln c|\right)$(the
latter value to be explained below). Noting from Eq. \ref{eq:14}
at $\varepsilon=\varepsilon_{loc}$ that the localization radius is
$r_{loc}=\hbar v_{\mathrm{F}}/\left(\Lambda\sqrt{c_{0}}\right)$,
this refers to $r_{loc}\ll\bar{r}=\hbar v_{\mathrm{F}}/\left(\Lambda\sqrt{\pi c}\right)$,
a 2D analogy to the metallization condition discussed in Introduction.
Thus, the characteristic concentration $c_{0}$ (in fact, an analog
to $n_{0}$ in Introduction) is bias tuned, and a tuned Anderson transition
on the $\varepsilon_{imp}$-band (merger of its mobility edges and
vanishing of band-like states \citep{and}) is expected at the bias
value $V_{A}$.

At growing bias to $V\gg c|U|$, the composite band structure can
be described with use of Eq. \ref{eq:17}. Then the renormalized dispersion
law $\varepsilon=\varepsilon_{imp}(\xi)$ follows from Eq. \ref{eq:11}
in an implicit form as:
\begin{eqnarray}
\frac{V^{2}}{4}-\varepsilon_{imp}^{2}(\xi) & = & \left(\frac{V^{2}}{4}-\varepsilon_{loc}^{2}\right)\nonumber \\
 & \times & \exp\left(\frac{c\Lambda^{2}}{\frac{V^{2}}{4}-\varepsilon_{imp}^{2}(\xi)+\xi^{2}}\right).\label{eq:18}
\end{eqnarray}
This equation permits analytic solutions near the edges of $\varepsilon_{imp}$-band.
Thus, the lower edge corresponds to $\xi\rightarrow0$:
\begin{equation}
\varepsilon_{g}\equiv\varepsilon_{imp}(0)\approx\frac{V}{2}-\frac{c\Lambda^{2}}{VW\left(c/c_{0}\right)},\label{eq:19}
\end{equation}
where $W\left(z\right)$ is the Lambert W-function \citep{corless}.
Its asymptotics, $W\left(z\gg1\right)\approx\ln\left(z/\ln z\right)$,
used in Eq. \ref{eq:19} provide a simpler function: 
\begin{equation}
\varepsilon_{g}\approx\frac{V}{2}+cU\left[1-\frac{\ln c}{\ln\left(cc_{0}\right)}\frac{V}{V_{A}}\right],\label{eq:20}
\end{equation}
replacing the above linear $\varepsilon_{g}$ dependence at $V\ll c|U|$
in the wider range of $V\ll V_{A}$. At yet higher bias, up to $V\sim V_{A}$,
the full Eq. \ref{eq:19} applies.

Expanding Eq. \ref{eq:18} at $\xi^{2}\ll V^{2}/4-\varepsilon_{g}^{2}$,
the long-wave dispersion law is obtained:

\begin{equation}
\varepsilon_{imp}(\xi)\approx\varepsilon_{g}+\left(1+\frac{V^{2}}{4c\Lambda^{2}}\right)^{-1}\frac{\xi^{2}}{2\varepsilon_{g}},\label{eq:21}
\end{equation}
indicating the $\varepsilon_{comp}$-behavior for $V\ll\sqrt{c}\Lambda$.
The further growth of $\varepsilon_{imp}(\xi)$ finally reaches the
short-wave asymptotics (at $\xi^{2}\gg c\Lambda^{2}$): 
\begin{equation}
\varepsilon_{imp}(\xi)\approx\varepsilon_{loc}-c\Lambda^{2}\frac{V/2-\varepsilon_{loc}}{\xi^{2}}.\label{eq:22}
\end{equation}
This defines its formal upper edge $\varepsilon_{f}=\varepsilon_{imp}(\Lambda)\approx\varepsilon_{loc}-c\left(V/2-\varepsilon_{loc}\right)$
\footnote{In fact, the real limits for $\varepsilon_{imp}$ band from the IRM
criterion, Eq. \ref{eq:12}, (the Mott's mobility edges) lie somewhat
deeper within this band, but as far as $c_{0}\ll c$ they do not sensibly
change the above estimate for $\Delta_{imp}$. %
}, and then the total width of impurity band:
\begin{equation}
\Lambda_{imp}\approx\varepsilon_{f}-\varepsilon_{g}\approx\frac{\Lambda^{2}}{V}\left(\frac{c}{W\left(c/c_{0}\right)}-c_{0}\right).\label{eq:23}
\end{equation}
Growing with $V$ from the initial value of $c|U|$, $\Lambda_{imp}$
by Eq. \ref{eq:23} would reach a maximum at some $V_{*}=F\left(c\right)\Lambda^{2}/|U|$
with the factor $F\left(c\right)$ varying from $\approx0.12$ to
$\approx0.22$ in the range of $10^{-4}<c<0.1$. However, such $V_{*}$
is already close to the critical value $V_{A}$ and hence to the impurity
band collapse, making this maximum meaningless.

The next step is to determine the lifetimes of the obtained quasiparticle
states, in order to establish the IRM limits for their existence.
For the quasiparticle with energy $\varepsilon=\varepsilon_{comp}\left(\xi\right)$,
we can consider its effective broadening:
\begin{eqnarray}
\Gamma_{comp}\left(\varepsilon\right) & = & \mathrm{Im}\,\frac{cU}{1-Ug_{1}\left(\varepsilon-cU\right)}\nonumber \\
 & \approx & \frac{\pi c}{2}\left(\frac{U}{\Lambda}\right)^{2}\frac{\left(V/2+\varepsilon-cU\right)^{2}}{\varepsilon-cU}.\label{eq:24}
\end{eqnarray}
Using it in the IRM criterion, Eq. \ref{eq:12}, we estimate the location
of the mobility edge $\varepsilon_{c}$ near the bottom of $\varepsilon_{imp}$:
$\varepsilon_{c}-\varepsilon_{g}\sim cU^{2}V/\Lambda^{2}$. It is
negligible beside the gap of $V$ between the $\varepsilon_{2}$-
and $\varepsilon_{imp}$-bands and the width $\Lambda_{imp}$ of $\varepsilon_{imp}$-band.
However, the broadening, Eq. \ref{eq:24}, at $\varepsilon\approx\varepsilon_{loc}$
much exceeds the formal gap $\approx V/2-\varepsilon_{loc}$ (here
exponentially small) between $\varepsilon_{imp}$- and $\varepsilon_{1}$-bands.
This permits to consider such gap and the very level $\varepsilon_{loc}$
non-existing and justifies the concept of a composite band in the
weak bias regime (see insets a,b in Fig. \ref{Fig2}). 

The overall electronic state of the doped system is determined by
the location of its Fermi level $\varepsilon_{\mathrm{F}}$ with respect
to the mobility edges. Supposing each impurity atom to supply one
carrier to the system and its undoped state to possess $\varepsilon_{\mathrm{F}}$=0,
its position at finite $c$ is found from the equation:
\begin{equation}
c=2\int_{\varepsilon_{g}}^{\varepsilon_{\mathrm{F}}}\rho\left(\varepsilon\right)d\varepsilon\label{eq:25}
\end{equation}
(including the spin factor 2). For the weak bias regime (or, in other
words, for $c_{0}\ll c$), one can use here the composite band DOS:
$\rho\left(\varepsilon\right)\approx\rho_{0}\left(\varepsilon+\varepsilon_{g}\right)$,
and obtain the bias dependent Fermi level within this band as: 
\begin{equation}
\varepsilon_{\mathrm{F}}\left(V\right)\approx\sqrt{c\Lambda^{2}+\varepsilon_{g}^{2}}.\label{eq:26}
\end{equation}
At low enough bias, $V\ll\sqrt{c}\Lambda$, it lies as high within
the $\varepsilon_{comp}$-band as $\varepsilon_{\mathrm{F}}\approx\sqrt{c}\Lambda$
(Fig. \ref{Fig2}) and defines a metallic behavior of the system.
This can be just compared to metallization of common semiconductors
at high enough doping ($c\gg c_{0}$). 

With growing bias, $\varepsilon_{\mathrm{F}}\left(V\right)$ gets
closer to the band's bottom $\varepsilon_{g}$ but its expected crossing
of a mobility edge and the system transition into insulating state
can be only reached after the decomposition of $\varepsilon_{imp}$
and $\varepsilon_{1}$ bands by means of an emerging localized range
around the impurity level $\varepsilon_{loc}$. From comparison of
Eqs. \ref{eq:15} and \ref{eq:24}, this is estimated to take place
at $c\sim c_{0}\ln^{2}\left(1/c_{0}\right)$. The related bias value
is high enough: 
\begin{equation}
V_{dec}\sim\frac{\Lambda^{2}}{2|UW_{-1}\left(-\sqrt{c}/2\right)|},\label{eq:27}
\end{equation}
including the lower branch $W_{-1}$ of the multivalued Lambert function
with asymptotics $W_{-1}\left(z\right)\approx\ln\left(z/\ln|z|\right)$
for $-1/{\rm e}<z<0$ \citep{corless}. However, this $V_{dec}$ is
yet well below the $\varepsilon_{imp}$-band collapse value $V_{A}$
(see inset c in Fig.\ref{Fig2}). Then, taking into account the T-matrix
contribution to DOS for the lower decomposed subband : 
\begin{equation}
\rho_{l}\left(\varepsilon\right)\approx\frac{\varepsilon}{\Lambda^{2}}-\frac{cc_{0}\Lambda^{2}}{2V^{2}\left(\varepsilon-\varepsilon_{loc}\right)},\label{eq:28}
\end{equation}
and using it in Eq. \ref{eq:25}, we find the condition that $\varepsilon_{\mathrm{F}}$
reaches the top of $\varepsilon_{imp}$: 
\begin{eqnarray}
c & = & 2\int_{\varepsilon_{g}}^{\varepsilon_{f}}\rho_{l}\left(\varepsilon\right)d\varepsilon\approx\frac{c}{W\left(c/c_{0}\right)}-c_{0}\nonumber \\
 & + & \frac{cc_{0}\Lambda^{2}}{2V^{2}}\ln\left(\frac{1}{c_{0}W\left(c/c_{0}\right)}-\frac{1}{c}\right).\label{eq:29}
\end{eqnarray}
Implicitly, Eq. \ref{eq:29} defines the bias value $V_{M}$ that
can be associated with the tuned MIT, provided this value be above
$V_{dec}$ so that the top of $\varepsilon_{imp}$ already pertain
to the localized range. The MIT bias value is estimated from Eq. \ref{eq:29}
as:

\begin{equation}
V_{M}\approx\frac{7\Lambda^{2}}{4|U|\ln\left(z_{M}/c\right)},\label{eq:30}
\end{equation}
where the factor in the logarithm depends on the perturbation parameter
as: $z_{M}\approx\left(3.35U/\Lambda\right)^{4}$, by a reasonable
empirical fit. Then, the numerical comparison between Eqs. \ref{eq:27}
and \ref{eq:30} shows that $V_{dec}$ in fact precedes $V_{M}$ for
all realistic $U\lesssim\Lambda$. But the sequence of tuned MIT and
Anderson transitions can be changed depending on the impurity parameters.
So, the $V_{A}\to V_{M}$ sequence for their above choice (as in Fig.
\ref{Fig2}) passes to $V_{M}\to V_{A}$ for $c=0.01$ and $|U|>0.71\Lambda$.
With the bias $V$ exceeding $V_{A}$, the impurity band does not
make sense already but there exists a well defined localized level,
Eq. \ref{eq:15}, whose width at $c\ll c_{0}$ estimated from the
GE pair term (see details in Appendix) becomes exponentially small:
$\Gamma_{loc}\sim c_{0}\left(\Lambda^{2}/V\right)\mathrm{e^{-\mathit{c_{\mathrm{0}}/c}}}$.

The above discussed tuned restructuring of spectrum can be summarized
as follows. At low bias, $V\ll V_{M}$, the system is metallized by
the impurity doping, with the Fermi level lying deep within the composite
$\varepsilon_{1}$+$\varepsilon_{imp}$-conduction band. The composite
band gets split into $\varepsilon_{1}$- and $\varepsilon_{imp}$-bands,
separated by the range of localized states around the impurity level
$\varepsilon_{loc}$, at bias reaching $V_{dec}$. At further growing
bias to $V_{M}$, the Fermi level meets the mobility edge above $\varepsilon_{loc}$
to produce MIT. After the Anderson transition occured at $V=V_{A}$,
a single $\varepsilon_{1}$ conduction (unoccupied) band is left in
the spectrum, the Fermi level staying fixed near $\varepsilon_{loc}$.
All the critical bias values, $V_{dec}$, $V_{M}$, and $V_{A}$,
can be reduced by choosing lower impurity concentration, but this
reduction is as slow as $\sim1/\ln\left(1/c\right)$ and simultaneously
the thermal stability level for MIT is reduced as $k_{\mathrm{B}}T_{max}\sim c\ln\left(1/c\right)\Lambda^{2}/|U|$. 

Now let us consider the alternative scenario, or the Anderson hybrid
model.

\section{Anderson hybrid model}

\begin{figure}[t]
\includegraphics[bb=140bp 190bp 800bp 620bp,scale=0.3]{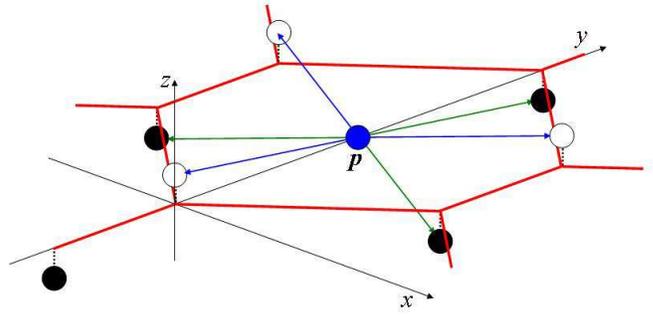}
\protect\caption{\label{Fig3}Anderson impurity at an interstitial position in silicene
lattice.}
\end{figure}

In the Anderson model, there are two perturbation parameters: the
on-site energy $\varepsilon_{0}$ for an electron at an impurity atom
and its modified hopping amplitude $\eta t$ (supposedly $\eta\lesssim1$)
to the nearest neighbor host sites. Such type of impurity perturbation
better corresponds to transition metal or rare-earth atoms, known
to produce deep levels in common Si. In silicene, these atoms predominantly
occupy interstitial positions, linked to both host sublattices (see
Fig. \ref{Fig3}). In the corresponding perturbation Hamiltonian:
\begin{eqnarray}
H_{A} & = & \sum_{{\bf p}}\left[\varepsilon_{0}\alpha_{\mathcal{\mathbf{{\bf p}}}}^{\text{\dag}}\alpha_{{\bf p}}\right.\nonumber \\
 & + & \left.\frac{1}{\sqrt{N}}\sum_{\mathbf{k}}\left({\rm e}^{i{\bf k}\cdot{\bf p}}\alpha_{\mathcal{\mathbf{p}}}^{\text{\dag}}\hat{\tau}_{\mathbf{k}}^{\dagger}\psi_{\mathbf{k}}+\mathrm{h.c.}\right)\right],\label{eq:31}
\end{eqnarray}
this linkage is presented through the 2-spinor $\hat{\tau}_{\mathbf{k}}^{\dagger}=\eta\left(t_{\mathbf{k}},t_{\mathbf{k}}^{*}\right)$.
The other difference of this model consists in the presence of independent
Fermi operators $\alpha_{{\bf p}}$ for an electron on impurity site
$\mathbf{p}$, besides the above considered host operators in the
$\psi_{\mathbf{k}}$ spinor. Subsequently, it generates more involved
GF structures (with respect to the sublattice indices): besides the
``host'' $\hat{G}({\bf k},\mathbf{k}')$ matrices by Eq. \ref{eq:2},
here also ``impurity'' $g_{{\bf p},{\bf p}'}=\langle\langle\alpha_{{\bf p}}|\alpha_{\mathcal{\mathbf{{\bf p}}}'}^{\text{\dag}}\rangle\rangle$
scalars and ``mixed'' $h_{{\bf k},{\bf p}}=\langle\langle\psi_{\mathbf{k}}|\alpha_{\mathcal{\mathbf{p}}}^{\text{\dag}}\rangle\rangle$
and $h_{{\bf p},{\bf k}}^{\dagger}=\langle\langle\alpha_{{\bf p}}|\psi_{\mathbf{k}}^{\dagger}\rangle\rangle$
spinors appear. 

Under the full Hamiltonian $H=H_{0}+H_{A}$, the equation of motion
for the ``host'' GF matrix gets modified from Eq. \ref{eq:7} to:

\begin{eqnarray}
\hat{G}({\bf k},\mathbf{k}') & =\delta_{{\bf k},\mathbf{k}'} & \hat{G}^{\left(0\right)}({\bf k})\nonumber \\
 & + & \frac{1}{\sqrt{N}}\sum_{{\bf p}}{\rm e}^{-i{\bf k}\cdot{\bf p}}\hat{G}^{\left(0\right)}({\bf k})\hat{\tau}_{\mathbf{k}}h_{{\bf p},{\bf k}'}^{\dagger},\label{eq:32}
\end{eqnarray}
and the respective equation for the ``mixed'' $h^{\dagger}$ spinor
in its right hand side reads as:
\begin{equation}
h_{{\bf p},{\bf k}'}^{\dagger}\left(\varepsilon-\varepsilon_{0}\right)=\frac{1}{\sqrt{N}}\sum_{{\bf k}''}{\rm e}^{i{\bf k}''\cdot{\bf p}}\hat{\tau}_{\mathbf{k''}}^{\dagger}\hat{G}({\bf k}'',{\bf k}').\label{eq:33}
\end{equation}
Then the specific solutions for all the above mentioned GF types follow
from consequent iterations of Eqs. \ref{eq:32} and \ref{eq:33}. 

The general strategy consists in separating, after each iteration
step, all the GF's in the right hand side that were already present
in the previous steps and doing next iteration for the resting ones,
in order to compose and then fully solve an equation for each GF.
The most important between them are the diagonal $\hat{G}({\bf k})$
and $g_{{\bf p}}\equiv g_{{\bf p},{\bf p}}$ that enter the total
DOS 
\begin{equation}
\rho\left(\varepsilon\right)=\frac{1}{2\pi N}\mathrm{\ Im}\,\left(\sum_{{\bf k}}\hat{G}_{{\bf k}}+\sum_{{\bf p}}g_{{\bf p}}\right),\label{eq:34}
\end{equation}
 in the generalization of Eq. \ref{eq:3}. Thus, a full solution for
$\hat{G}({\bf k})$ follows from Eq. \ref{eq:32} (at ${\bf k}=\mathbf{k}'$),
after separating the same $\hat{G}({\bf k})$ in the right hand side
of Eq. \ref{eq:33} for $h_{{\bf p},{\bf k}}^{\dagger}$ and applying
again Eq. \ref{eq:32} to all $\hat{G}({\bf k}',\mathbf{k})$ with
${\bf k}'\neq\mathbf{k}$ there. At this next iteration, a similar
separation of $h_{{\bf p},{\bf k}}^{\dagger}$ is also done, giving
rise to a respective full solution for $h_{{\bf p},{\bf k}}^{\dagger}$,
and so on. The result, formally analogous to Eq. \ref{eq:8}:

\begin{equation}
\hat{G}({\bf k})=\left\{ \left[\hat{G}^{\left(0\right)}({\bf k})\right]^{-1}-\hat{\varSigma}_{{\bf k}}\right\} ^{-1},\label{eq:35}
\end{equation}
includes the self-energy matrix in the GE form, similar to Eq. \ref{eq:9}:
$\hat{\varSigma}_{{\bf k}}=c\hat{T}_{{\bf k}}\left(1+cB_{{\bf k}}+\dots\right)$.
But here the T-matrix term: 
\begin{equation}
c\hat{T}_{{\bf k}}=\frac{\eta^{2}\xi^{2}}{1+2\eta^{2}}\frac{1}{N}\sum_{{\bf p}}\frac{1+\hat{\sigma}_{+}{\rm e}^{2i\varphi}+\hat{\sigma}_{-}{\rm e}^{-2i\varphi}}{\varepsilon-\varepsilon_{imp}-i\Gamma_{imp}-\Sigma_{{\bf p}}},\label{eq:36}
\end{equation}
has an important difference from the Lifshitz model form, Eq. \ref{eq:9},
in its momentum dependence, both on the radial variable $\xi\equiv\xi_{\mathbf{k}}$
and on the angular argument $\varphi\equiv\varphi_{\mathbf{k-K}}$.
It also includes the single impurity level $\varepsilon_{imp}=\varepsilon_{0}/\left(1+2\eta^{2}\right)$
(reduced by its coupling to the host) with its imaginary part $\Gamma_{imp}=\mathrm{Im}\,\sum_{{\bf k}}\hat{\tau}_{\mathbf{k}}^{\dagger}\hat{G}_{{\bf k}}\hat{\tau}_{\mathbf{k}}$
and the ``impurity'' scalar self-energy:

\begin{eqnarray}
\Sigma_{\boldsymbol{\mathbf{p}}} & = & \sum_{\mathbf{{\bf p}}'\neq\mathcal{\mathbf{{\bf p}}}}A_{\mathcal{\mathbf{{\bf p}}}-\mathcal{\mathbf{{\bf p}}}'}\left[F_{\mathcal{\mathbf{{\bf p}}}'-\mathcal{\mathbf{{\bf p}}}}\right.\nonumber \\
 & + & \left.\sum_{\mathbf{{\bf p}}''\neq\mathcal{\mathbf{{\bf p\mathit{,}p'}}}}A_{\mathcal{\mathbf{{\bf p}}}'-\mathcal{\mathbf{{\bf p}}}''}\left(F_{\mathcal{\mathbf{{\bf p}}}''-\mathcal{\mathbf{{\bf p}}}}+\ldots\right)\right].\label{eq:37}
\end{eqnarray}
Here the scalar functions 
\begin{eqnarray*}
F_{{\bf \mathcal{\mathbf{{\bf p}}}-\mathcal{\mathbf{{\bf p}}}'}} & = & \frac{1}{N}\sum_{{\bf k}}{\rm e}^{i{\bf k}\cdot{\bf \left(\mathcal{\mathbf{{\bf p}}}-\mathcal{\mathbf{{\bf p}}}'\right)}}\hat{\tau}_{\mathbf{k}}^{\dagger}\hat{G}_{{\bf k}}\hat{\tau}_{\mathbf{k}},\\
\mathrm{A{}_{\mathcal{\mathbf{{\bf p}}}-\mathcal{\mathbf{{\bf p}}}'}} & = & \frac{F_{{\bf \mathcal{\mathbf{{\bf p}}}-\mathcal{\mathbf{{\bf p}}}'}}}{\varepsilon-\varepsilon_{imp}-i\Gamma_{imp}-\Sigma_{{\bf p}'}},
\end{eqnarray*}
describe the effects of indirect interactions between impurity centers.
The latter $A_{\mathcal{\mathbf{{\bf p}}}-\mathcal{\mathbf{{\bf p}}}'}$
functions also define the GE terms of the ``host'' self-energy $\hat{\varSigma}_{{\bf k}}$,
along the same formal structure as in Eq. \ref{eq:10}, while the
GE structure for the scalar $\Sigma_{\boldsymbol{\mathbf{p}}}$ in
Eq. \ref{eq:37} is notably different. In this way, the solution for
the diagonal ``impurity'' GF follows as: 
\begin{equation}
g_{{\bf p}}=\frac{1}{\left(1+2\eta^{2}\right)\left(\varepsilon-\varepsilon_{imp}-i\Gamma_{imp}-\Sigma_{{\bf p}}\right)},\label{eq:38}
\end{equation}
and can be then used in Eq. \ref{eq:34}. A specific feature of $\Sigma_{\boldsymbol{\mathbf{p}}}$
is the random statistical distribution of its values due to random
$\mathcal{\mathbf{{\bf p}}}'$ positions around given ${\bf p}$,
with the standard deviation $\sigma=\sqrt{\overline{\Sigma_{\boldsymbol{\mathbf{p}}}^{2}}-\overline{\Sigma_{\boldsymbol{\mathbf{p}}}}^{2}}$.
Since a finite range of inter-impurity interactions, this deviation
does not vanish in the thermodynamical limit $N\to\infty$, unlike
that for $\hat{\varSigma}_{{\bf k}}$ (known as the self-averaging
property \citep{lifgrepas}). As to the mean self-energy $\overline{\Sigma_{\boldsymbol{\mathbf{p}}}}$,
it can be simply included into the definition of impurity level $\varepsilon_{imp}$,
so understood in what follows. 

\begin{figure}[h]
\includegraphics[scale=0.5]{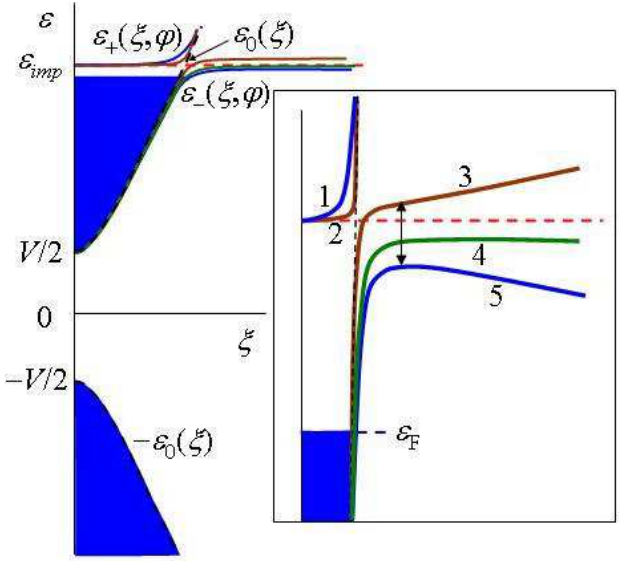}
\protect\caption{\label{Fig4}Silicene dispersion laws near their splitting by the
impurity level $\varepsilon_{imp}$ (red dashed line) show an in-plane
anisotropy. Inset resolves them for particular directions: 1 for $\varepsilon_{+}\left(\xi,\pi\right)$,
2 for $\varepsilon_{+}\left(\xi,0\right)$, 3 for $\varepsilon_{-}\left(\xi,0\right)$,
4 for $-\left(\xi,\pi/2\right)$, 5 for $\varepsilon_{-}\left(\xi,\pi\right)$.
The impurity parameters are chosen as: $\varepsilon_{imp}=0.1\Lambda$,
$\tilde{c}=0.005$ and the bias $V=0.4\varepsilon_{imp}$. The arrows
indicate the splitting range near the impurity level $\varepsilon_{imp}$.}
\end{figure}

In this model, we present the basic secular determinant as $\mathrm{det}\,\left(\hat{G}_{\mathbf{k}}\right)^{-1}=\mathrm{det}\,\left(\hat{G}_{\mathbf{k}}^{\left(0\right)}\right)^{-1}+\Sigma_{{\bf k}}$,
with the scalar self-energy: 
\begin{eqnarray}
\Sigma_{{\bf k}} & = & \frac{\eta^{2}\xi^{2}\left(\varepsilon+\xi\cos\varphi\right)}{1+2\eta^{2}}\nonumber \\
 & \times & \frac{1}{N}\sum_{{\bf p}}\frac{1}{\varepsilon-\varepsilon_{imp}-i\Gamma_{imp}-\Sigma_{{\bf p}}},\label{eq:39}
\end{eqnarray}
Then the general Eq. \ref{eq:11} (written at the T-matrix level in
neglect of $\Gamma_{imp}$ and $\Sigma_{{\bf p}}$) reads:

\begin{equation}
\varepsilon^{2}=\varepsilon_{0}^{2}\left(\xi\right)+2\tilde{c}\xi^{2}\frac{\varepsilon+\xi\cos\varphi}{\varepsilon-\varepsilon_{imp}}\label{eq:40}
\end{equation}
where the reduced impurity concentration $\tilde{c}=c\eta^{2}/\left(1+2\eta^{2}\right)$
measures the quasiparticle weight transfer from impurity to band states.
Eq. \ref{eq:40} defines the dispersion laws for ``host'' quasiparticles
that are quite close to the non-perturbed $\varepsilon_{0}\left(\xi\right)$
except for the $\varepsilon_{imp}$ vicinity of $\sim\tilde{c}\varepsilon_{imp}$
width where the splitting of two subbands is mainly developed. Within
that splitting range, both subbands strongly deviate from $\varepsilon_{0}\left(\xi\right)$
and display a sensible in-plane anisotropy: $\varepsilon_{\pm}\left(\xi,\varphi\right)$
(Fig. \ref{Fig4}), unlike the isotropically split subbands in Fig.
\ref{Fig2}. Physically, this anisotropy reflects the breakdown of
local inversion symmetry for an impurity interstice at applied field
bias.

Other difference from the Lifshitz model is in the possibility that,
at varying bias $V$, the impurity level $\varepsilon_{imp}$ can
be crossed by the band edge $V/2$. Lastly, the presence of $\xi^{2}$
factor in the T-matrix, Eq. \ref{eq:36}, leads to vanishing damping
for long-wave quasiparticles, so that the mobility edges should correspond
to shorter wavelengths (to be defined below). This implies that Bloch-like
states with such short wavelengths would not make sense at energies
close enough to the impurity level. 

As a result, the composition of electronic spectrum in the Anderson
model is more complicated than in the Lifshitz model. Here we have
generally up to three subbands of the states by electrons on host
sites: the two split $\varepsilon_{\pm}\left(\xi,\varphi\right)$
subbands and the almost non-perturbed $-\varepsilon_{0}\left(\xi\right)$
(valence) subband, together with a subband of the states on impurity
interstices (Eq. \ref{eq:38}). The important separation between band-like
and localized states can be established from the following principle.
A given energy $\varepsilon$ pertains to the range of band-like states
by virtue of those solutions of Eq. \ref{eq:39} that satisfy the
IRM criterion, and if no such solutions exist, this energy pertains
to the range of localized states by virtue of the related solutions
of Eq. \ref{eq:38}. All these states contribute to the total DOS
with their particular weights defined by the residues of corresponding
poles of diagonal GF's in Eq. \ref{eq:34}.

Following this principle, one can apply the IRM criterion, Eq. \ref{eq:12},
for almost isotropic band-like states beyond the splitting range,
$|\varepsilon-\varepsilon_{imp}|\gg\tilde{c}\varepsilon_{imp}$, but
will need its more complete form:

\begin{equation}
{\bf k}\cdot\nabla_{{\bf k}}\varepsilon_{\pm}\left(\xi,\varphi\right)\gg\Gamma_{\pm}\left(\xi,\varphi\right),\label{eq:41}
\end{equation}
within this range. Expecting the most important mobility threshold
to be located below $\varepsilon_{imp}$, we focus on the relevant
damping term $\Gamma_{-}\left(\xi,\varphi\right)=\mathrm{Im}\,\Sigma_{{\bf k}}|\varepsilon-\varepsilon_{imp}|/|\varepsilon^{2}-V^{2}/4|$.
There are several additive contributions to this term, due to $\Gamma_{imp}$
and $\Sigma_{{\bf p}}$ in the denominator of $\hat{T}_{{\bf k}}$,
Eq. \ref{eq:36}, and to the GE terms such as $B_{{\bf k}}$. A more
detailed analysis shows that the dominating contribution comes from
$\Sigma_{{\bf p}}$, expressing the decay rates of band quasiparticles
into the randomly distributed localized levels $\varepsilon_{imp}+\Sigma_{{\bf p}}$.
This contribution is already self-averaging and its average involves
the probability distribution function $P\left(\Sigma_{{\bf p}}\right)$.
Since $\Sigma_{{\bf p}}$ by Eq. \ref{eq:37} is a sum of a great
number of independent random values like $A_{{\bf p-p'}}F_{{\bf p'-p}}$,
this probability distribution should take a normal form: 
\[
P\left(\Sigma_{{\bf p}}\right)=\frac{1}{\sqrt{2\pi}\sigma}\mathrm{\exp}\left[-\frac{\left(\Sigma_{{\bf p}}-\overline{\Sigma_{{\bf p}}}\right)^{2}}{2\sigma^{2}}\right],
\]
that readily implies:
\begin{eqnarray}
\mathrm{Im}\,\Sigma_{{\bf k}} & = & 2\tilde{c}\xi^{2}\left(\varepsilon+\xi\cos\varphi\right)\mathrm{Im}\,\overline{\left(\varepsilon-\varepsilon_{imp}-\Sigma_{{\bf p}}+\overline{\Sigma_{{\bf p}}}\right)^{-1}}\nonumber \\
 & = & \sqrt{2\pi}\tilde{c}\xi^{2}\frac{\varepsilon+\xi\cos\varphi}{\sigma}\mathrm{\exp}\left[-\frac{\left(\varepsilon-\varepsilon_{imp}\right)^{2}}{2\sigma^{2}}\right].\label{eq:42}
\end{eqnarray}
Now, to evaluate the standard deviation $\sigma$, we restrict $\Sigma_{{\bf p}}$,
Eq. \ref{eq:37}, to its first term and approximate the interaction
function by using the non-perturbed GF (see details in Appendix):
\begin{eqnarray}
F_{{\bf \mathcal{\mathbf{{\bf n}}}}} & \approx & \frac{1}{N}\sum_{{\bf k}}{\rm e}^{i{\bf k}\cdot\mathcal{\mathbf{{\bf p}}}}\hat{\tau}_{\mathbf{k}}^{\dagger}\hat{G}_{{\bf k}}^{\left(0\right)}\hat{\tau}_{\mathbf{k}}\approx\frac{2c_{0}}{\pi}\nonumber \\
 & \times & \left[\varepsilon K_{0}\left(\frac{n}{r_{\varepsilon}}\right)-\Lambda\frac{n\cos\theta}{r_{\varepsilon}}K_{1}\left(\frac{n}{r_{\varepsilon}}\right)\right].\label{eq:43}
\end{eqnarray}
Since within the relevant energy range for this case, $V/2<\varepsilon$,
the argument of McDonald functions turns to be imaginary, they can
be expressed through the 1st and 2nd kind Bessel functions \citep{Abst}:
$K_{0}\left(ix\right)=\frac{\pi}{2}\left[Y_{0}\left(x\right)+iJ_{0}\left(x\right)\right]$
and $K_{1}\left(ix\right)=\frac{\pi}{2}\left[-J_{1}\left(x\right)+iY_{1}\left(x\right)\right]$.
Notice the presence of $p$-wave anisotropy by $\cos\theta=n_{x}/n$
in Eq. \ref{eq:43}, similar to that of the dispersion law, Eq. \ref{eq:39}.
Next we obtain (see Appendix):

\begin{eqnarray}
\sigma^{2} & \approx & c\sum_{{\bf n}\neq0}\left(A_{{\bf n}}F_{{\bf -n}}\right)^{2}\nonumber \\
 & \sim & \frac{\left(c_{0}\varepsilon_{imp}\right)^{4}}{\left(\varepsilon-\varepsilon_{imp}\right)^{2}}\frac{c}{c_{cr}}\ln\frac{1}{c_{cr}},\label{eq:44}
\end{eqnarray}
presenting an energy-dependent $\sigma\left(\varepsilon\right)$ that
grows at approach to the single impurity level $\varepsilon_{imp}$,
as it can be expected for such resonance interactions. In Eqs. \ref{eq:43},
\ref{eq:44}, $c_{0}=|\varepsilon_{imp}^{2}-V^{2}/4|/\Lambda^{2}$
generalizes the definition in Eq. \ref{eq:15} for the locations of
band edge $V/2$ either above and below $\varepsilon_{imp}$, and
the meaning of $c_{cr}=\varepsilon_{imp}^{2}/\Lambda^{2}$ is explained
in what follows. Using the result by Eq. \ref{eq:44} in Eq. \ref{eq:42},
we conclude that the relevant damping term $\Gamma_{-}\left(\xi,\varphi\right)$
very steeply shoots up from exponentially low to values as high as
$\sim\left(\tilde{c}/c_{0}\right)\left(\varepsilon+\xi\cos\varphi\right)\sim\varepsilon_{imp}$
when reaching the condition $|\varepsilon-\varepsilon_{imp}|\sim\sigma\left(\varepsilon\right)$
or: 
\begin{equation}
|\varepsilon-\varepsilon_{imp}|\sim\left(\frac{c}{c_{cr}}\ln\frac{1}{c_{cr}}\right)^{1/4}c_{0}\varepsilon_{imp}.\label{eq:45}
\end{equation}
So Eq. \ref{eq:45} gives just an estimate for the distance from $\varepsilon_{imp}$
to the mobility edge $\varepsilon_{c}$. We notice that, for $c$
close to $c_{cr}$ and $c_{0}$, this distance exceeds the width $\tilde{c}\varepsilon_{imp}$
of the splitting range that justifies the above usage of the non-perturbed
spectrum in Eq. \ref{eq:43}. Also, the estimate by Eq. \ref{eq:45}
defines the above referred minimum admitted wavelength of band quasiparticles
with energies near $\varepsilon_{imp}$: $\lambda_{min}\sim a/\sqrt{c_{0}}$.

The principal practical issue of tuned metal-insulator transition
is resolved by comparing the mobility edge $\varepsilon_{c}$ and
the Fermi level $\varepsilon_{\mathrm{F}}$ whose initial position
at zero bias is below $\varepsilon_{imp}$ if the impurity concentration
is smaller of the above defined critical value: $c<c_{cr}$. 
\begin{widetext}

\begin{figure}[t]

\includegraphics[clip,scale=0.6]{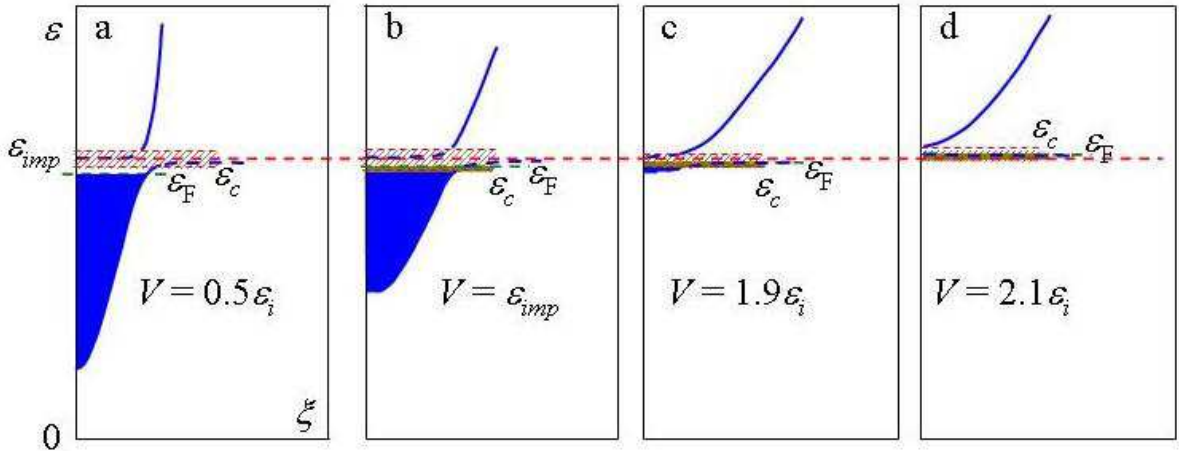}

\protect\caption{\label{Fig5}Dispersion curves for silicene with Anderson impurities
as in Fig. \ref{Fig4} at growing bias $V$ that moves the bottom
of the conduction band towards the impurity level $\varepsilon_{imp}$.
MIT occurs at $V_{M}\approx0.84\varepsilon_{imp}$ (in between a and
b panels) and the Anderson transition at $V_{A}\approx2\varepsilon_{imp}$
(in between c and d panels). }

\end{figure}

\end{widetext}Expecting
the crossing of $\varepsilon_{\mathrm{F}}$ with $\varepsilon_{c}$
at growing bias to occur outside the splitting range, we can safely
estimate $\varepsilon_{\mathrm{F}}$ with use of the non-perturbed
DOS, Eq. \ref{eq:5}, in Eq. \ref{eq:25} to result in:
\begin{equation}
\varepsilon_{\mathrm{F}}\left(V\right)\approx\sqrt{c\Lambda^{2}+V^{2}/4},\label{eq:46}
\end{equation}
instead of Eq. \ref{eq:26}. Then the MIT bias $V_{M}$,
when the Fermi level crosses the mobility edge, is found for the Anderson
model as: 
\begin{equation}
V_{M}\approx2\varepsilon_{imp}\sqrt{1-\frac{c}{c_{cr}}\left(1+2c^{1/4}c_{cr}^{3/4}\ln^{1/4}\frac{1}{c_{cr}}\right)},\label{eq:47}
\end{equation}
that is slightly below the critical value $V_{cr}=2\varepsilon_{imp}\sqrt{1-c/c_{cr}}$,
when the Fermi level reaches $\varepsilon_{imp}$ (as shown in Fig.
\ref{Fig5}). With further growing bias, $V>V_{cr}$, the Fermi level
stays fixed near $\varepsilon_{imp}$ while the Anderson transition
for the $\varepsilon_{-}\left(\xi,\varphi\right)$ subband takes place
when the mobility edge is attained by the band edge $V/2$. This corresponds
to $V_{A}\approx2\varepsilon_{imp}$. 

Finally, at $V>V_{A}$, the impurity level $\varepsilon_{imp}$ stays
below the bottom of the almost unperturbed main band (as in Fig. \ref{Fig5}d),
then $\Sigma_{\boldsymbol{\mathbf{p}}}$ and so the broadening of
$\varepsilon_{imp}$ becomes exponentially small $\sim c_{0}\Lambda\mathrm{e^{-\mathit{c_{\mathrm{0}}/c}}}$
(at $c_{0}\gg c$) by virtue of similar decay of $F_{{\bf n}}$ in
Eq. \ref{eq:43} and there are almost unperturbed subbands $\pm\varepsilon_{0}\left(\xi\right)$
in the spectrum (alike the above case of the Lifshitz model).

\section{Discussion and conclusions}

\label{sec:conc}The above presented considerations of electronic
spectrum in biased and doped silicene within the frameworks of two
models for impurity perturbation show a variety of restructuring processes
in this spectrum with different dynamics for its particular subbands,
derived from both the initial host subbands and from the impurity
levels. The main difference between the two models is in the location
of impurity energy level. In the Lifshitz model, it is $\varepsilon_{loc}$,
Eq. \ref{eq:15}, which closely follows the biased edge of one of
the main subbands while in the Anderson model it is the bias independent
$\varepsilon_{imp}$, Eq. \ref{eq:33}, that can be crossed by the
biased main band edge. This determines the difference in spectrum
transformations with growing bias $V$. 

\begin{figure}[t]
\includegraphics[scale=0.4]{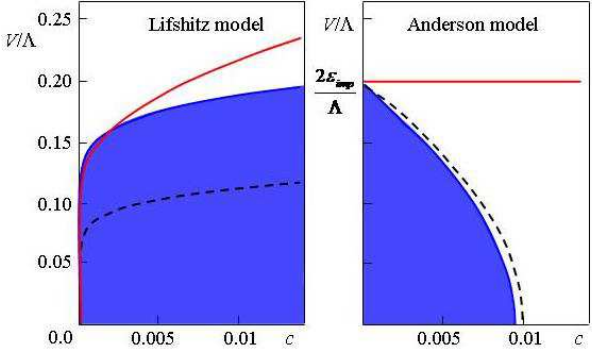}

\protect\caption{\label{Fig6}Phase diagrams of electronic states in biased doped silicene
in the variables ``impurity concentration-electric bias'' for two
models of impurity perturbation. The blue areas correspond to metallic
phases, separated by the Mott MIT lines $V_{M}$ (dark blue) from
insulating phases (white areas), Anderson transitions (collapse of
impurity band) are shown by the red lines, dashed lines indicate decoupling
of the impurity band from the main band, $V_{dec}$ (in the Lifshitz
model), or complete filling of all the states up to the impurity level
$\varepsilon_{imp}$ (in the Anderson model). }
\end{figure}

In the Lifshitz model, the initial metallic state at $V\ll V_{M}$
corresponds to the Fermi level $\varepsilon_{\mathrm{F}}$ well above
$\varepsilon_{loc}$ and the tuned MIT is realized through its dropping
down to the mobility edge that emerges near $\varepsilon_{loc}$ after
the impurity subband gets decoupled from its neighbor main subband
at $V>V_{dec}.$ This process develops rather slowly with growing
bias and requires the stronger critical level $V_{M}$ the higher
impurity concentration $c$ is present in the sample. In contrast,
the Anderson model provides a possibility for initial $\varepsilon_{\mathrm{F}}$
to be positioned below $\varepsilon_{imp}$ and to reach the mobility
edge with growing bias, the sooner the higher $c$ is chosen. These
scenarios can be suitably presented in the form of phase diagrams
in terms of the relevant variables ``impurity concentration-electric
bias'' (Fig. \ref{Fig6}).

These diagrams for the two models clearly display the above mentioned
difference in electronic phase dynamics. In the Lifshitz model, the
critical value as a function of impurity concentration, $V_{M}\left(c\right)$,
grows (logarithmically slowly) (Fig. \ref{Fig6}a), while in the Anderson
model this function is rapidly decreasing from its initial value $V_{M}\left(0\right)=2\varepsilon_{imp}$
(Fig. \ref{Fig6}b). Also, there is a notable difference in behavior
of other phase boundary in this case which defines the Anderson transition
in the collapsing impurity band. In the Lifshitz model, $V_{A}\left(c\right)$
grows in a similar way to $V_{M}\left(c\right)$ and, depending on
the perturbation parameter $U$, a crossing of these two can take
place. By contrast, $V_{A}\left(c\right)$ in the Anderson model is
practically constant: $V_{A}\left(0\right)\approx2\varepsilon_{imp}$. 

It is readily seen from Eq. \ref{eq:47} that a considerable reduction
of the MIT bias can be reached by driving the impurity concentration
$c$ close enough to $c_{cr}$. This is an essential advantage of
the Anderson model scenario compared to that of the Lifshitz model,
also taking into account that growing $c$ simultaneously improves
the thermal stability of tuned MIT. The other practical advantage
here is in a much higher steepness $s_{M}=d\left(\varepsilon_{c}-\varepsilon_{\mathrm{F}}\right)/dV$
of this transition. This is seen from the comparison of corresponding
values: $V_{M}\approx0.34\Lambda$, $s_{M}\approx0.24$ for the Lifshitz
model in Fig. \ref{Fig3} and $V_{M}\approx0.08\Lambda$, $s_{M}\approx0.42$
for the Anderson model in Fig. \ref{Fig5}. The latter advantage is
even more enforced by the fact that the damping of Fermi quasiparticles
(defining the Drude resistivity of metal) at $V\to V_{M}$ varies
slowly in the Lifshitz model, as in Eq. \ref{eq:24}, but grows exponentially
in the Anderson model, as in Eq. \ref{eq:42}, enabling here an extremely
strong variation of the doped system resistivity near the tuned MIT. 

Summarizing, the doped and biased silicene presents a suitable opportunity
for realization of practical electronic devices with tunable electric
resistivity over a very broad scale, from normal metallic to fully
insulating (possibly accompanied by respective optical, thermal, etc.
effects), under rselatively weak bias. This regime can be optimized
by a proper choice of impurity atoms, their location within the crystalline
structure, and their concentration. The comparative analysis of two
common models for impurity perturbation on the host electronic spectrum
indicates the Anderson hybrid model (adequate for transition or rare
earth impurities in silicene) to be more promising for such purpose.
Comparing the present system to the other known material with tunable
gap, the bigraphene, where similar doping effects were recently considered
\citep{PSL}, an advantage of the silicene host is seen in the simpler
structure of its electronic spectrum. Experimental checks on the proposed
regimes of doping and tunable phase transitions could better determine
the field for future studies and probably open some new possibilities
in this direction.

\section*{Acknowledgements}

Y.G.P. is grateful for the support from Portuguese FCT by the Project
No. PTDC/FIS/120055/2010. The work of V.M.L. was partly supported
by the Special Program for Fundamental Research of the Division of
Physics and Astronomy of the National Academy of Sciences of Ukraine.

\appendix

\section*{Appendix}

Quasimomentum sums over the Brillouin zone (BZ) commonly result in
certain analytic functions of other relevant arguments (such as position
vectors, energy, etc.), and their calculation is done by passing from
sum to integral

\begin{equation}
\frac{1}{N}\sum_{{\bf k}}f_{{\bf k}}=\frac{1}{v_{BZ}}\int_{BZ}f_{{\bf k}}d{\bf k},\label{eq:48}
\end{equation}
where $v_{BZ}$ is the BZ volume. In the present case, the 2D BZ consists
of two equilateral triangles with side $\sqrt{3}K$, each centered
in a nodal point, and the above integration of a function $f_{{\bf k}}$
that decays fast enough away from the nodal points can be approximately
done in the radial and angular variables $\xi$, $\varphi$ over the
circle of radius $\Lambda$:
\begin{equation}
\frac{1}{v_{BZ}}\int_{BZ}f_{{\bf k}}d{\bf k}\approx\frac{1}{\pi\Lambda^{2}}\int_{0}^{\Lambda}\xi d\xi\int_{0}^{2\pi}d\varphi f\left(\xi,\varphi\right).\label{eq:49}
\end{equation}
Among all the functions that have no such decay and so do not admit
such approximation, we distinguish the important case of plane wave,
$f_{{\bf k}}={\rm e}^{i{\bf k}\cdot\mathcal{\mathbf{{\bf n}}}}$,
providing an exact result:

\begin{equation}
\frac{1}{N}\sum_{{\bf k}}{\rm e}^{i{\bf k}\cdot\mathcal{\mathbf{{\bf n}}}}=\delta\left(\mathcal{\mathbf{{\bf n}}}\right),\label{eq:50}
\end{equation}
where the discrete Dirac delta is $\delta\left(\mathcal{\mathbf{{\bf n}}}\right)=0$
for any $\mathcal{\mathbf{{\bf n}}}$ joining two lattice sites (or
two interstices) and $\delta\left(0\right)=1$. 

Now we apply these techniques to the calculation of the basic interaction
functions $A_{{\bf n}}$ and $F_{{\bf n}}$ in the approximation of
non-perturbed spectrum. Starting from the definition in the Lifshitz
model, Eq. \ref{eq:10}, we obtain (at $\varepsilon^{2}<V^{2}/4$):
\begin{eqnarray*}
A_{j,{\bf n}} & \approx & \frac{T_{j}\left(\varepsilon\pm V/2\right)}{N}\sum_{{\bf k}}\frac{\mathrm{e}^{i{\bf k}\cdot{\bf n}}}{\varepsilon^{2}-V^{2}/4-\xi^{2}}\\
 & \approx & \frac{T_{j}\left(2\varepsilon\pm V\right)}{\Lambda^{2}}\int_{0}^{\infty}\xi d\xi\frac{J_{0}\left(\xi p/\hbar v_{\mathrm{F}}\right)}{\varepsilon^{2}-V^{2}/4-\xi^{2}},\\
 & = & \frac{T_{j}\left(\varepsilon\pm V/2\right)}{\Lambda^{2}}K_{0}\left(n/r_{\varepsilon}\right),
\end{eqnarray*}
the result in Eq. \ref{eq:14}. In particular, its asymptotics at
$n\gg r_{\varepsilon}$ defines the broadening of localized impurity
level at $c\ll c_{0}$, by the criterion $c\mathrm{Im}\, B_{1}\sim1$,
where:
\begin{eqnarray*}
\mathrm{Im}\, B_{1} & = & \mathrm{Im}\,\sum_{{\bf n}\neq0}\frac{1}{1-A_{1,{\bf n}}^{2}}\approx\frac{\pi}{a^{2}}\int_{a}^{\infty}\delta(1-A_{1,{\bf r}}^{2})rdr\\
 & \approx & \frac{\pi r_{\varepsilon}^{2}}{a^{2}}\ln\frac{c_{0}\Lambda^{2}}{V\left(\varepsilon-\varepsilon_{loc}\right)}\approx\frac{\pi}{c_{0}}\ln\frac{c_{0}\Lambda^{2}}{V\left(\varepsilon-\varepsilon_{loc}\right)}.
\end{eqnarray*}
Therefore the above criterion takes place at $\left|\varepsilon-\varepsilon_{loc}\right|\sim\left(c_{0}\Lambda^{2}/V\right)\mathrm{e}^{-c_{0}/c}$,
as indicated after Eq. \ref{eq:30}.

For the Anderson model, with the definition by Eq. \ref{eq:37}, we
have (at $\varepsilon^{2}>V^{2}/4$): 
\begin{eqnarray}
F_{{\bf n}} & \approx & \frac{2}{N}\sum_{{\bf k}}{\rm e}^{i{\bf k}\cdot\mathcal{\mathbf{{\bf n}}}}\xi^{2}\frac{\varepsilon+\xi\cos\varphi}{\varepsilon^{2}-V^{2}/4-\xi^{2}}\nonumber \\
 & =- & \frac{2}{N}\sum_{{\bf k}}{\rm e}^{i{\bf k}\cdot\mathcal{\mathbf{{\bf n}}}}\left(\varepsilon+\xi\cos\varphi\right)\nonumber \\
 & + & \frac{2\left(\varepsilon^{2}-V^{2}/4\right)}{N}\sum_{{\bf k}}\frac{{\rm e}^{i{\bf k}\cdot\mathcal{\mathbf{{\bf n}}}}\left(\varepsilon+\xi\cos\varphi\right)}{\varepsilon^{2}-V^{2}/4-\xi^{2}}.\label{eq:51}
\end{eqnarray}
Then, applying Eq. \ref{eq:50} to the first sum in the right hand
side of Eq. \ref{eq:51} gives two terms proportional to $\delta\left(\mathcal{\mathbf{{\bf n}}}\right)$
and $d\delta\left(\mathcal{\mathbf{{\bf n}}}\right)/dn_{x}$, that
vanish for any $\mathcal{\mathbf{{\bf n}}}\neq0$. The second sum
is treated by means of Eq. \ref{eq:49} as follows:
\begin{eqnarray*}
\frac{2}{N}\sum_{{\bf k}}{\rm e}^{i{\bf k}\cdot\mathcal{\mathbf{{\bf n}}}}\frac{\varepsilon+\xi\cos\varphi}{\varepsilon^{2}-V^{2}/4-\xi^{2}}\quad\quad\quad\quad\quad\\
\approx\frac{2}{\pi\Lambda^{2}}\int_{0}^{\Lambda}\frac{\xi d\xi}{\varepsilon^{2}-V^{2}/4-\xi^{2}}\quad\quad\quad\quad\\
\times\left[\varepsilon J_{0}\left(\frac{\xi n}{\hbar v_{\mathrm{F}}}\right)+\xi J_{1}\left(\frac{\xi n}{\hbar v_{\mathrm{F}}}\right)\cos\theta\right],
\end{eqnarray*}
and, after extending the upper integration limit to infinity, the
exact formulas can be used:
\begin{eqnarray}
\int_{0}^{\infty}\frac{xJ_{0}\left(x\right)dx}{b^{2}-x^{2}} & = & \frac{i\pi}{2}H_{0}^{\left(2\right)}\left(b\right),\nonumber \\
\int_{0}^{\infty}\frac{x^{2}J_{1}\left(x\right)dx}{b^{2}-x^{2}} & = & \frac{i\pi}{2}bH_{1}^{\left(2\right)}\left(b\right),\label{eq:52}
\end{eqnarray}
with the 2nd kind Hankel functions $H_{l}^{\left(2\right)}\left(x\right)$
\citep{Abst} and $b=n/|r_{\varepsilon}|$. This leads finally to
the result of Eq. \ref{eq:43}.

Now let us consider the standard deviation of the random scalar self-energy
$\Sigma_{{\bf p}}$:
\begin{equation}
\sigma^{2}=\overline{\Sigma_{{\bf p}}^{2}}-\overline{\Sigma_{{\bf p}}}^{2},\label{eq:53}
\end{equation}
restricting the definition of $\Sigma_{{\bf p}}$ in Eq. \ref{eq:37}
to its first term, of the lowest linear order in impurity concentration.
Then we present:
\begin{eqnarray}
\Sigma_{{\bf p}} & \approx & \left(\varepsilon-\varepsilon_{imp}\right)^{-1}\sum_{{\bf n}\neq0}c_{{\bf n}}F_{{\bf n}}F_{-{\bf n}},\nonumber \\
\Sigma_{{\bf p}}^{2} & \approx & \left(\varepsilon-\varepsilon_{imp}\right)^{-2}\sum_{{\bf n},{\bf n}'\neq0}c_{{\bf n}}c_{{\bf n}'}F_{{\bf n}}F_{-{\bf n}}F_{{\bf n}'}F_{-{\bf n}'},\label{eq:54}
\end{eqnarray}
where the random numbers $c_{{\bf n}}$ of impurity occupation at
${\bf n}$th interstice take the values 1 with probability $c$ and
0 with probability $1-c$ and the non-renormalized denominators $\varepsilon-\varepsilon_{imp}$
correspond to the adopted precision to the lowest order in $c$. Using
this in Eq. \ref{eq:53} and taking into account that $\overline{c_{{\bf n}}}=c$,
$\overline{c_{{\bf n}}c_{{\bf n}'}}=c^{2}$ at ${\bf n}\neq{\bf n}'$
and $\overline{c_{{\bf n}}^{2}}=c$, we obtain:
\begin{equation}
\sigma^{2}=c\left(1-c\right)\left(\varepsilon-\varepsilon_{imp}\right)^{-2}\sum_{{\bf n}\neq0}\left(F_{{\bf n}}F_{-{\bf n}}\right)^{2}.
\end{equation}
The decisive point in the evaluation of this sum over interstices
is that the products $F_{{\bf n}}F_{-{\bf n}}$ do not include Fourier
components with coinciding quasimomenta (see after Eq. \ref{eq:10}).
Using the inverse orthogonality relation to Eq. \ref{eq:50}, $N^{-1}\sum_{{\bf n}}{\rm e}^{i{\bf k}\cdot\mathcal{\mathbf{{\bf n}}}}=\delta({\bf k})$,
we present the relevant sum as:
\begin{eqnarray*}
\sum_{{\bf n}\neq0}\left(F_{{\bf n}}F_{-{\bf n}}\right)^{2} & = & -F_{0}^{4}+S,
\end{eqnarray*}
where $F_{0}^{4}\approx\left(c_{0}\varepsilon\right)^{4}$ is prevailed
by 
\[
S=\frac{1}{N^{3}}\sum_{{\bf k}_{1},{\bf k}_{2},{\bf k}_{3}}f_{{\bf k}_{1}}f_{{\bf k}_{2}}f_{{\bf k}_{3}}f_{{\bf k}_{1}+{\bf k}_{2}-{\bf k}_{3}},
\]
 with 
\[
f_{{\bf k}}=2\left(\varepsilon^{2}-V^{2}/4\right)\frac{\varepsilon+\xi\cos\varphi}{\varepsilon^{2}-V^{2}/4-\xi^{2}}.
\]
 This triple sum in ${\bf k}_{j}$ is dominated by its real part,
mainly due to the short-wave contributions by $\xi\gg\sqrt{\varepsilon^{2}-V^{2}/4}\approx\sqrt{c_{0}}\Lambda$,
and it can be estimated by the triple integral in $\xi_{j}=\hbar v_{{\rm F}}k_{j}$:
\begin{eqnarray}
S & \sim & \left(2c_{0}\varepsilon_{imp}\right)^{4}\Lambda^{2}\int_{\varepsilon_{imp}}^{\Lambda}\frac{d\xi_{1}}{\xi_{1}}\int_{\varepsilon_{imp}}^{\Lambda}\frac{d\xi_{2}}{\xi_{2}}\nonumber \\
 & \times & \int_{\varepsilon_{imp}}^{\Lambda}\frac{d\xi_{3}}{\xi_{3}\left(\xi_{1}^{2}+\xi_{2}^{2}+\xi_{3}^{2}\right)}\sim\frac{\left(c_{0}\varepsilon_{imp}\right)^{4}}{c_{cr}}\ln\frac{1}{c_{cr}}
\end{eqnarray}
(the $\varphi$-oscillating terms in $f_{{\bf k}}$ are not important
for this result). This readily leads to the expressions in Eqs. \ref{eq:44}
and \ref{eq:45}. 

Finally, the broadening of localized impurity level at $V/2>\varepsilon_{imp}$
is obtained in the same way as shown above for the case of the Lifshitz
model at $c\ll c_{0}$, with only difference for the pre-exponential
factor $c_{0}\Lambda$, here resulting from the dominant contribution
by the second term in Eq. \ref{eq:43}.

\end{document}